\documentclass[aps,twocolumn,amsmath,amssymb]{revtex4}

\usepackage{graphicx}
\usepackage{dcolumn}
\usepackage{bm}
\usepackage{amssymb}
\usepackage{amsmath}
\usepackage{epsfig}
\usepackage{subfigure}

\begin{document}


\title{The complex channel networks of bone structure}

\author{Luciano da Fontoura Costa}\email{luciano@if.sc.usp.br} \author{Matheus Palhares Viana}
 \affiliation{Instituto de F\'{\i}sica de S\~{a}o Carlos, Universidade
 de S\~{a}o Paulo, Av. Trabalhador S\~{a}o Carlense 400, Caixa Postal
 369, CEP 13560-970, S\~{a}o Carlos, S\~ao Paulo, Brazil}

\author{Marcelo Em\'{i}lio Beletti}
\affiliation{
Instituto de Ci\^{e}ncias Biom\'{e}dicas, Universidade Federal de Uberl\^{a}ndia\\
Rua Par\'{a}, 1720, CEP 38400-902, Uberl\^{a}ndia, Minas Gerais, Brasil
}

\date{18th Dec 2004}

\begin{abstract}

Bone structure in mammals involves a complex network of channels
(Havers and Volkmann channels) required to nourish the bone marrow
cells.  This work describes how three-dimensional reconstructions of
such systems can be obtained and represented in terms of complex
networks.  Three important findings are reported: (i) the fact that
the channel branching density resembles a power law implies the
existence of distribution hubs; (ii) the conditional node degree
density indicates a clear tendency of connection between nodes with
degrees 2 and 4; and (iii) the application of the recently introduced
concept of hierarchical clustering coefficient allows the
identification of typical scales of channel redistribution.  A series
of important biological insights is drawn and discussed.

\end{abstract}


\maketitle

Containing two types of channels, namely Havers and Volkmann channels,
the bones of mammals are pervaded by a network of intercommunicating
channels required to maintain the bone living cells~\cite{1}, which is
done by blood vessels distributed inside those channels.  Although
playing a decisive role in animal development, diseases, and bone
reconstitution, the topology of the bone channel network remains a
subject of investigation.  Because of its generality, the concept of
complex networks~\cite{2,3} stands out as a natural and comprehensive
means to represent, characterize and model the bone channel system.
By associating the branching points of channels to the nodes of the
network, and representing the connections between such spots in terms
of arcs between the network nodes, it is possible to represent the
bone channel structure in terms of a complex network. Topological
measurements obtained from such networks --- especially the node
degree density~\cite{4}, the conditional node density~\cite{5}, and
the \textit{hierarchical clustering coefficient}~\cite{6,7} --- can
then be used to identify important underlying features of the analyzed
channel system.

The bone used as a biological model in this study is the humerus of
the adult cat, collected from necropsy cats in the Sector of Pathology
at the Veterinary Hospital of the Federal University of Uberlandia,
Brazil. After dissection of the skin and muscles, a ring 0.5 cm wide
was removed from the central portion of both right and left humerus of
each animal.  Such rings were fixed in 10\% formalin for at least 48
hours. After setting, these samples were decalcified in 4\% nitric
acid per 30 days, embedded in paraffin according to a classic
histological technique, and sectioned in microtome in order to obtain
200 serial sections with 5 $\mu$m thickness.  These sections were then
stained by the Schmorl procedure, and digital images were acquired by
using an Olympus Triocular BX40 microscope connected to an Oly-200
camera interfaced to a PC-compatible computer through digital plate
Datatranslation 3135.  After the digitalization procedure, we
extracted the regions of interest through the segmentation~\cite{8} of both
Havers and Volkmann channels.  

\begin{figure*}
	\subfigure[]{\includegraphics[scale=0.4]{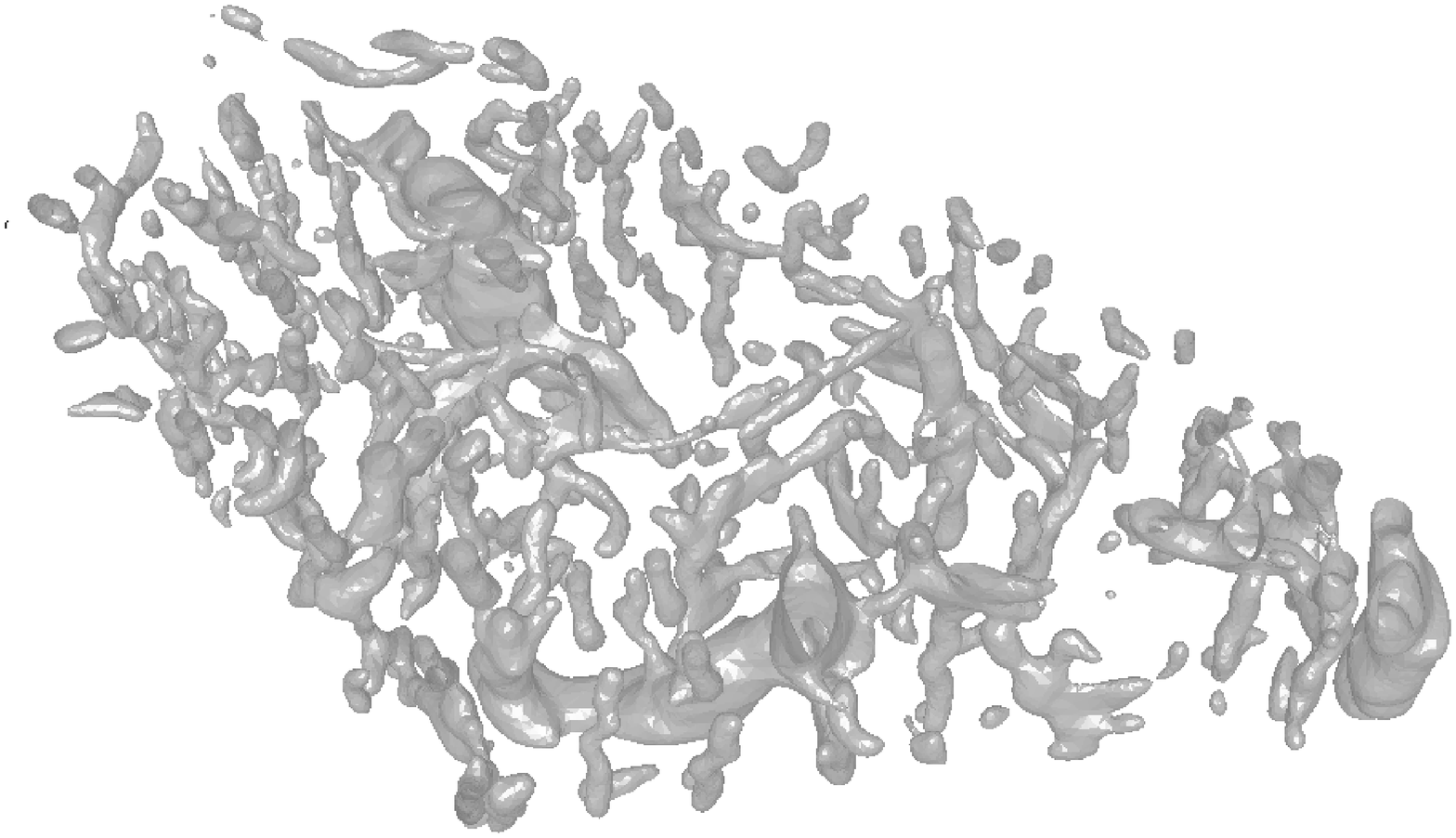}}
	\subfigure[]{\includegraphics[scale=0.25]{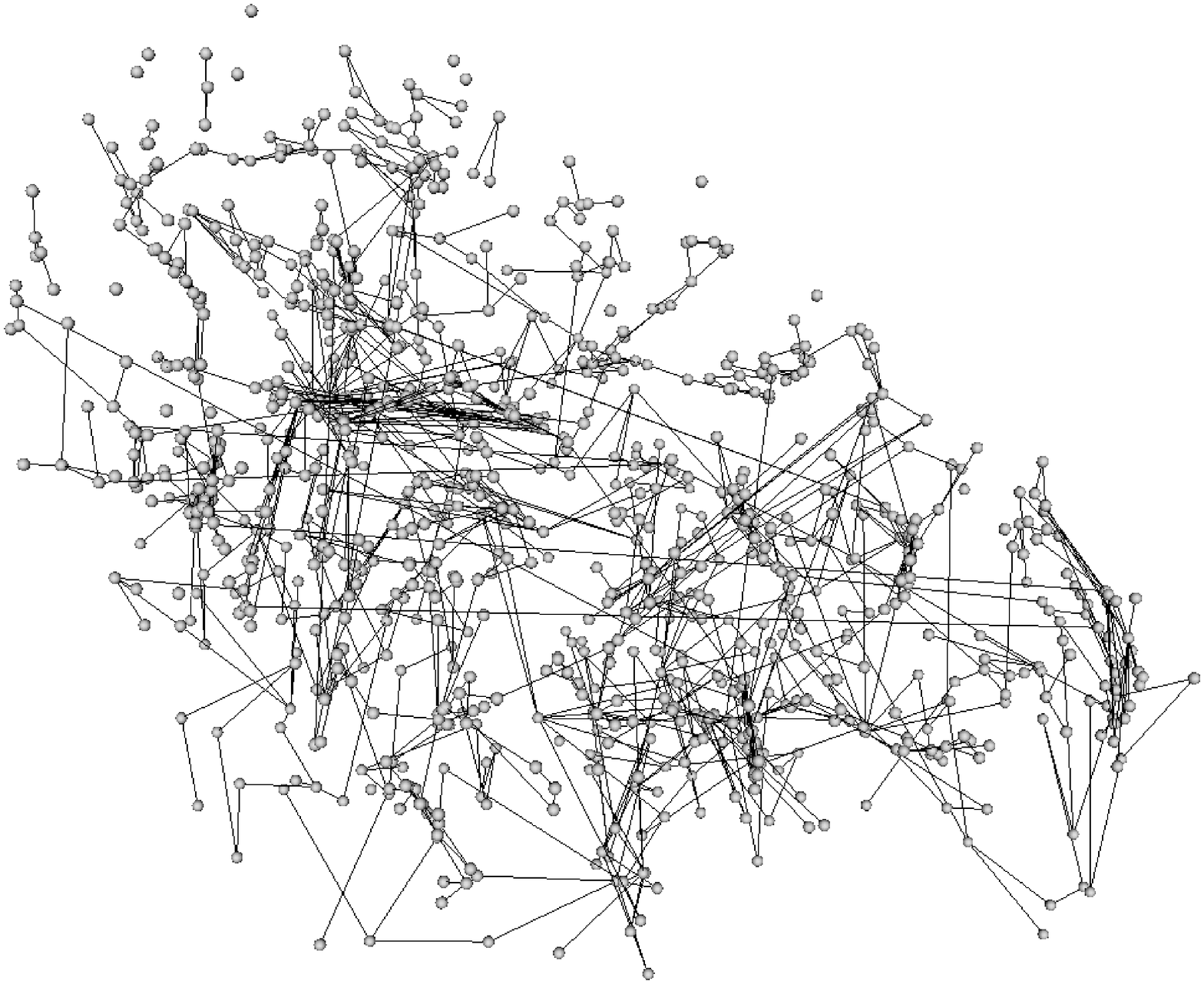}}
\caption{The three-dimensional reconstruction of the on-fourth of the considered bone channel network (a) and the respectively complex
network (b) obtained by projecting the node coordinates along the $z-$axis.}
\label{f1}
\end{figure*}

\begin{figure*}
	\subfigure[]{\includegraphics[scale=0.5]{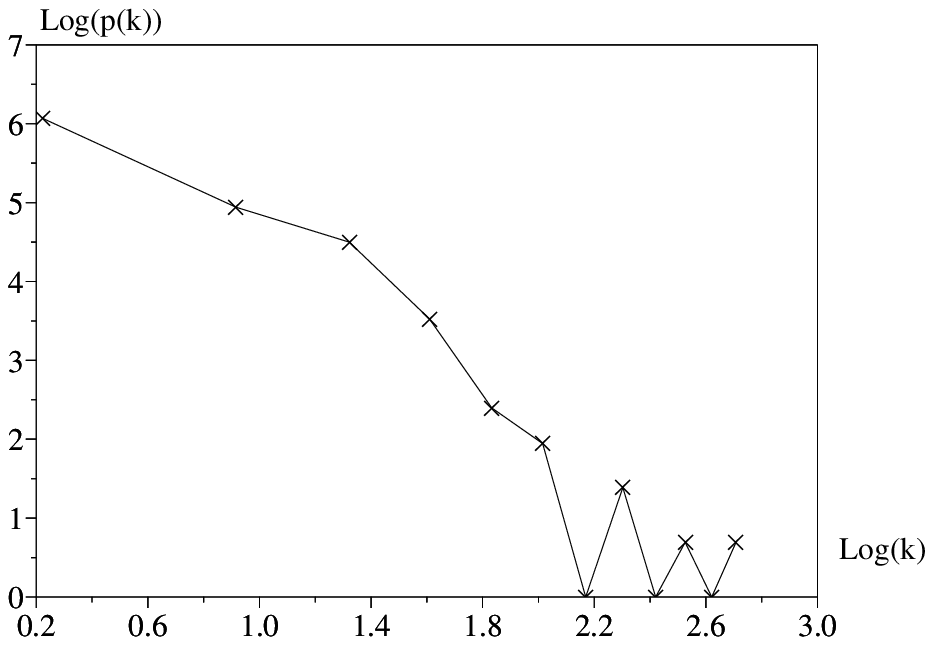}}
	\subfigure[]{\includegraphics[scale=0.55]{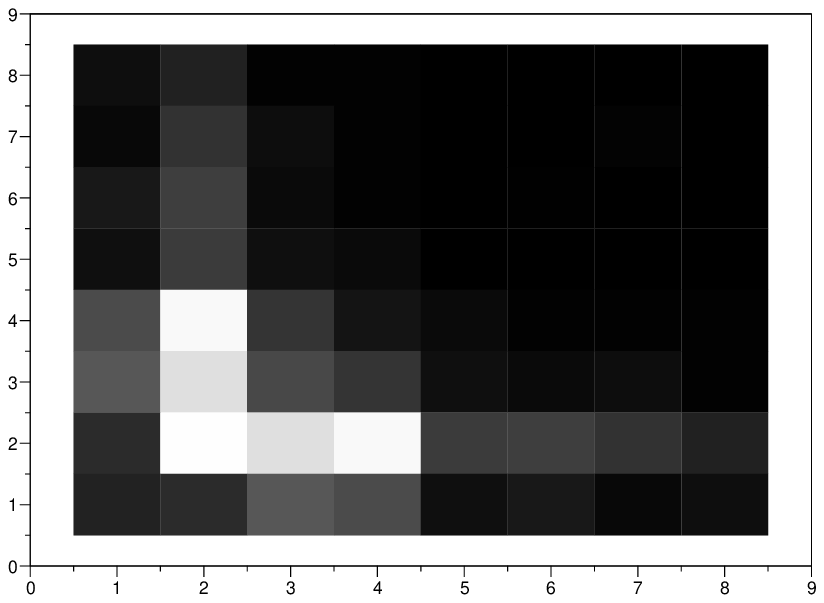}}
	\subfigure[]{\includegraphics[scale=0.5]{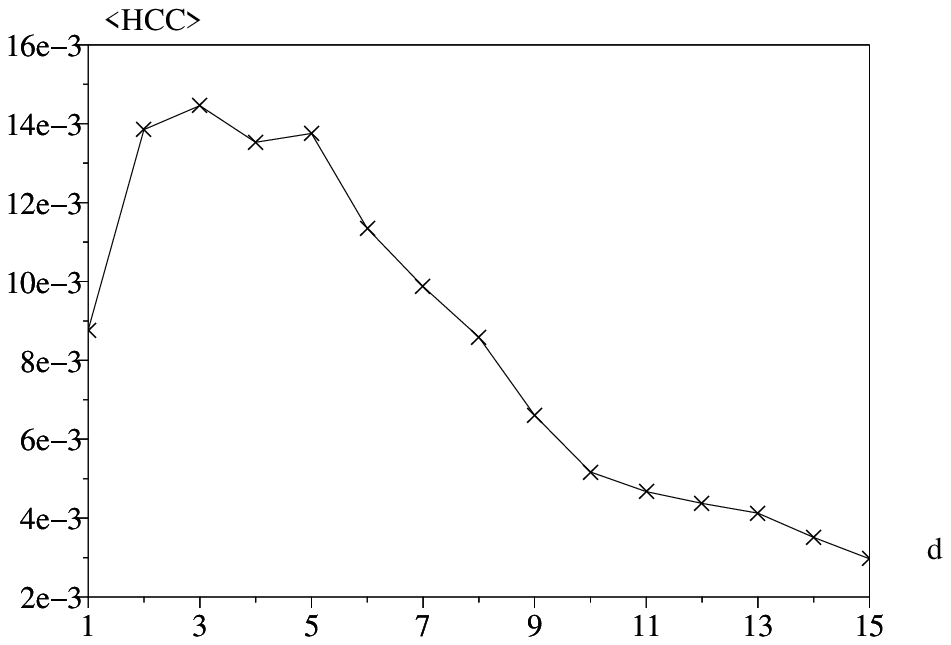}}
	\caption{The dilog plot of the node degree density, shown in (a),
suggests a scale-free network model and indicates the presence of
distribution hubs.  The conditional node degree density is presented
in (b), expressing the predominance of connections between nodes with
degrees 2 and 4. The average values of the hierarchical clustering
coefficients, shown in (c), is characterized by a plateu extending
from $d=2$ to 5.}
	\label{f2}
\end{figure*}

The images with final size of approximately 700$\times$480 pixels were
converted into PNG format and used to obtain a three-dimensional
reconstruction of the studied structure, which was performed by using
the vtkPNGReader class in the Visualization Toolkit for C++.
Figure~\ref{f1}(a) shows the three-dimensional reconstruction~\cite{9,10} of
part of the analyzed bone structure (one fourth).  The detection of
nodes of the graph and respective connections was performed manually,
yielding an unweighted, non-oriented complex network, which was
represented in terms of its $N \times N$ adjacency matrix $K$, such
that the existence of a connection between any two nodes $i$ and $j$
is represented by making $K(i,j)=K(j,i)=1$.  The vertices of
``V''-like channels were understood to correspond to branching spots
with missing connections, giving rise to nodes with degree 2.
Figure~\ref{f1}(b) shows the geographic complex network (the nodes
have positions in $R^2$) obtained from the structure in
Figure~\ref{f1}(a) by projecting the node coordinates along the
$z-$ axis.  The degree of a node $i$ is defined as the number of edges
connected to that node, which can be calculated as $k(i) =
\sum_{q=1}^{N} K(i,q)$.  Figure~\ref{f2}(a) shows the dilog plot
of the node degree density obtained for the bone considered in the
present study, which is characterized by 988 nodes and 1120 links.

The obtained density profile indicates that the bone channel network
resembles a scale free model~\cite{11}, which is characterized by the
existence of \emph{hubs}, i.e. nodes with particularly high degrees
which act as concentrators of connections.  Such a finding rules out
the possibility that the channel system could be organized as a
regular network or a tree, both characterized by similar node degrees
throughout. The conditional node density of the bone channel structure
is shown in Figure~\ref{f2}(b).  It is clear from such a density that
nodes with degree 2 tend to connect with nodes with degree 4, which
may be related to the fact that the circulation of the arteries inside
the channels follows an Eulerian network, which is characterized by
nodes with even degree~\cite{12}.

In order to further understand the topological structure of the
channel system, we considered the recently introduced concept of
hierarchical clustering coefficient.  Take a generic
node in the network, represented by $i$.  Let $H(i,d)$ be the set of
nodes which are at distance $d$ from node $i$ (the distance between
two nodes $i$ and $j$ corresponds to the number of edges along the
shortest path between those nodes).  Observe that the consideration of
subsequent values of $d$ defines a hierarchy around the reference node
$i$.  Now, the hierarchical clustering coefficient of node $i$ at
distance $d$ can be defined as follows

\begin{equation}
  HCC(i,d) = 2 \frac{E(H(i,d))}{||H(i,d)|| (||H(i,d)||-1)}
\end{equation}

where $E(S)$ is the number of edges between the nodes in a set $S$ and
$||S||$ is the number of elements (or cardinality) in that set.
Observe that the traditional clustering coefficient is obtained by
making $d=1$ in the above equation.  It can be verified that the
hierarchical clustering coefficient of node $i$ at distance $d$
quantifies the connectivity between those network nodes which are at
distance $d$ from $i$.  Observe also that $0 \leq HCC(i,d) \leq 1$,
with the null value indicating total absence of connections between
the nodes in $H(i,d)$, while the unit value indicates that each node
in $H(i,d)$ is connected to all other nodes in that set, except itself
(loops are not taken into account in this work).  Of particular
interest is the fact that higher clustering coefficient at a specific
distance $d$ implies the presence of many cycles composed of $2d+1$
edges.

Figure~\ref{f2}(c) shows the average value $\left< HCC \right>$
(considering each network node) of the hierarchical clustering
coefficient for $d=1$ to $15$.  Starting at 0.008 for $d=1$, the
values of $\left< HCC \right>$ then define a plateau extending from
$d=2$ to 5 and then falls steadily.  Such a result provides important
insights into the topological organization of the channel system.
First, the low clustering coefficient observed for $d=1$ expresses the
fact that the neighbors of each node are weakly connected one
another. This indicates that the channel structure does not involve
cycles at such a scale.  At the same time, the relatively higher
values of hierarchical clustering coefficients observed from $d=2$ to
5 implies the existence of several cycles of 5 to 11 edges and provides
further indication that the anlysed channel system is not a tree.

Generally, the arterial vascularization of the majority of tissues and
organs takes place through dichotomic bifurcation, i.e. an artery
bifurcates into two arteries, which in turn bifurcate into four arteries, and
so on.  The obtained experimental results provide cogent indication
that the vascularization of the bone structure does not follow
such a general rule.  A possible explanation for the deviation from
the traditional tree and the presence of cycles could be related to
different constraints imposed by specific tissues over structural
reconstruction after eventual artery obstruction, which can deprive
whole regions of proper blood flow.  While in tissues other than bones
the organism is known to initiate a neovascularization intended to
provide bypasses to region irrigation, such schemes are virtually
impossible in the short term in bones, which would demand a complex
and long reorganization of both Havers and Volkmann channels containing
the arteries (recall that the bone matrix is rigid).  Therefore, it
seems that the particularly intense connections between nodes from the
second to the fifth hierarchical levels, as indicated by the obtained
hierarchical clustering coefficients, represents a possible
redistribution system intended to immediately cope with eventual
artery obstruction.

All in all, this work has described how concepts from
three-dimensional reconstruction and image analysis, used jointly with
state-of-the-art complex network theory, can be effectively applied in
order to quantify certain topological features of biological
interconnecting systems, particularly the channel distribution network
in mammals bones. We have verified that such structures involve the
presence of distribution hubs, present a tendency for node connections
with degree 2 and 4, and contain cycles with typical
lengths ranging from 5 to 11 edges.  Important biological implications
of such topological features have been identified and
discussed. Further related investigations include the possibility to
quantify the width and length of each channel in order to infer
additional insights about the bone channel system by using weighted
and geographic complex networks.

\begin{acknowledgments}
Luciano da F. Costa thanks HFSP RGP39/2002, FAPESP (proc. 99/12765-2)
and CNPq (proc. 3082231/03-1) for financial support.
\end{acknowledgments}

\end{document}